\documentclass[doublecol]{epl2}

\usepackage{subfigure}
\usepackage{amsmath}
\usepackage{cite}
\usepackage{enumitem}
\setenumerate{noitemsep,topsep=0pt,parsep=0pt,partopsep=0pt}

\title{Node-weighted interacting network measures improve the representation of real-world complex systems}
\shorttitle{Node-weighted interacting network measures}

\author{M. Wiedermann\inst{1,2}{\thanks{E-mail: \email{marcw@physik.hu-berlin.de}}} \and J.F. Donges\inst{1,2,3}{\thanks{E-mail: \email{donges@pik-potsdam.de}}} \and J. Heitzig\inst{1} \and J. Kurths\inst{1,2,4}}
\shortauthor{M. Wiedermann \etal}

\institute{
  \inst{1} Potsdam Institute for Climate Impact Research - P.O. Box 60\,12\,03, 14412 Potsdam, Germany\\
  \inst{2} Department of Physics, Humboldt University - Newtonstr. 15, 12489 Berlin, Germany\\
  \inst{3} Stockholm Resilience Centre, Stockholm University - Kr\"aftriket 2B, 114\,19 Stockholm, Sweden\\
  \inst{4} Institute for Complex Systems and Mathematical Biology, University of Aberdeen - Aberdeen AB24 3FX, United Kingdom
}

\abstract{
Network theory provides a rich toolbox consisting of methods, measures, and models for studying the structure and dynamics of complex systems found in nature, society, or technology. Recently, it has been pointed out that many real-world complex systems are more adequately mapped by networks of interacting or interdependent networks, e.g., a power grid showing interdependency with a communication network. Additionally, in many real-world situations it is reasonable to include node weights into complex network statistics to reflect the varying size or importance of subsystems that are represented by nodes in the network of interest. E.g., nodes can represent vastly different surface area in climate networks, volume in brain networks or economic capacity in trade networks. In this letter, combining both ideas, we derive a novel class of statistical measures for analysing the structure of networks of interacting networks with heterogeneous node weights. Using a prototypical spatial network model, we show that the newly introduced node-weighted interacting network measures indeed provide an improved representation of the underlying system's properties as compared to their unweighted analogues. We apply our method to study the complex network structure of cross-boundary trade between European Union (EU) and non-EU countries finding that it provides important information on trade balance and economic robustness.
}

\begin{document}

\maketitle


\section{Introduction}

Complex network theory has been shown to be a powerful tool for analysing the structure and function of many complex systems in nature, society, and technology. Various kinds of measures have been defined recently, mostly based on counting nodes, paths, links or triangles in a network \cite{boccaletti2006complex, newman2010networks, cohen2010complex}. The field of application is wide-spread, e.g., considering social \cite{newman2005measure}, trade \cite{baskaran2011}, biological \cite{achard2006resilient, zhou2006hierarchical}, communication \cite{capocci2006preferential} and climate networks \cite{tsonis2006networks, Yamasaki2008, donges2009backbone, donges2011investigating, heitzig2012node}.
When applying network theory to real-world networks, it is often not sufficient to describe the underlying complex system by an isolated network. Instead, a network of interacting networks may provide an improved representation as was shown, e.g., in an analysis of the interdependency between the Internet network and the Italian power grid during a blackout in 2008~\cite{buldyrev2010catastrophic}. In general, interdependent networks behave much differently from isolated ones in terms of robustness to random failure, expected network properties \cite{parshani2010interdependent,parshani2010intersimilarity} as well as synchronisation behaviour \cite{Li2009synchronization}. 
In order to quantify the structure of interdependent networks, the recently introduced interacting networks approach compromises a set of cross-network measures in analogy to the canonical network measures designed for isolated networks \cite{schultz2010coupled, donges2011investigating, feldhoff2012geometric}. The method has been applied successfully for analysing the dynamical structure of the lower atmosphere by constructing coupled climate networks from pairs of geopotential height fields at different isobaric surfaces \cite{schultz2010coupled, donges2011investigating}. 

When applying network theory to real-world problems, nodes need not all bear the same importance for the network's properties, but there may be nodes having a strong impact on the network's topology that is not reflected by link properties alone. E.g., nodes can represent vastly different surface area in climate networks \cite{donges2009backbone, donges2011investigating, heitzig2012node, tsonis2006networks}, volume in brain networks \cite{achard2006resilient, zhou2006hierarchical} or economic capacity in trade networks \cite{garlaschelli2005structure, hidalgo2007product}. Therefore standard network measures treating all nodes the same may not represent appropriately all structural properties of the underlying complex system (in the following referred to as the domain of interest) \cite{Bialonski2010}. To take this into account, it is useful to assign an individual weight to every node according to the share of the whole domain of interest that is represented by that node. The concept of \emph{node splitting invariance (n.s.i.)} yields a set of measures allowing for a precise estimation of network characteristics when dealing with inhomogeneous node weights~\cite{heitzig2012node}.

The main aim of this work is to combine the frameworks of n.s.i.\ and interacting network measures in order to provide a general tool for investigating interaction structure within a network of networks with heterogeneous node weights (Fig \ref{fig:IN_scheme}). Starting from the cross-network measures \cite{donges2011investigating}, we introduce a general construction mechanism to derive \emph{n.s.i.\ cross-network measures}. Subsequently we apply these measures to a global model network with a structure prototypical of spatially embedded networks of networks and validate the results for n.s.i.\ cross-degree $k_v^{j*}$ and n.s.i.\ cross-clustering coefficient $C_v^{j*}$. To demonstrate the potential of our approach, we carry out an analysis of a trade network divided into two subnetworks, EU and non-EU countries, where nodes represent individual countries weighted according to their Gross Domestic Product (GDP) as a proxy of economic power.

\section{Methods}

\begin{figure}
    \centering
    \includegraphics[width=0.8\columnwidth]{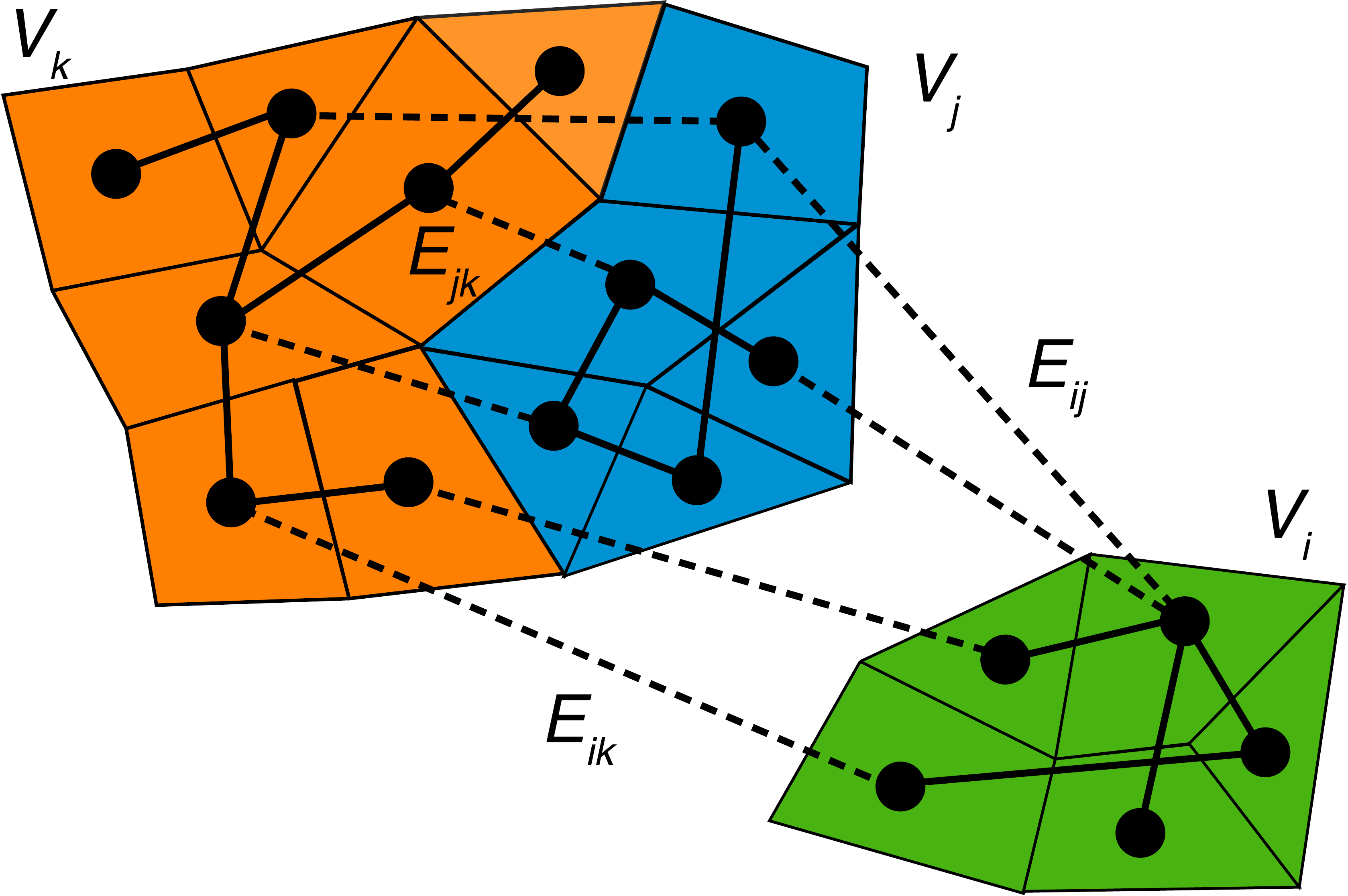}
    \caption{(Colour online) A set of three subnetworks $G_i=(V_i,E_{ii})$ (green), $G_j=(V_j,E_{jj})$ (blue) and $G_k=(V_k,E_{kk})$ (orange) connected via both internal (solid lines) and cross- (dashed lines) links. Each node $v$ represents a share of the whole domain of interest and is therefore provided with an individual weight $w_v$. The node sets $V_i$ and $V_j$ share a common boundary in some generalised space, while $V_k$ remains separated.}
    \label{fig:IN_scheme}
\end{figure}

We consider a network $G=(V,E)$ with a given set of nodes $V$, links $E$ and the number of nodes $N=|V|$. As we identify every node $v\in V$ with a natural number $p\leq N$, the network $G$ is represented by its adjacency matrix A with $A_{pq} = 1~\text{if}\  \{p,q\}\in E,\ A_{pq} =  0~\text{if}\ \{p,q\} \not\in E$.

Let this network be divided into $M\ge 2$ subnetworks $G_i=(V_i, E_{ii}), i=1,2,...,M$ with subsets of nodes $V_i$, such that $V=\bigcup_iV_i$ with $V_i\cap V_j=\emptyset$ for $i\neq j$. The set of links $E$ then splits into two types of links. There are sets of internal links $E_{ii}$ that connect nodes $v\in V_i$ within a subnetwork $G_i$ and \emph{cross-links} $E_{ij}$ that connect nodes $v\in V_i$ with nodes $q\in V_j$ in the subnetworks $G_i$ and $G_j$, so that $E = \bigcup_{i,j}E_{ij}$ (Fig.~\ref{fig:IN_scheme}). Notice that a network might be divided into several subnetworks by splitting a continuous domain of interest into smaller sub-domains, e.g.,\ when dividing a landmass into adjacent countries forming subnetworks ($G_i$ and $G_j$ in Fig.~\ref{fig:IN_scheme}). However, it might also occur that the two sub-domains are separated, since they display different observables or layers, such as a power grid with power plants as nodes and transmission lines as links which show interdependency with a communication network, where servers are represented by nodes and links indicate wires between computers ($G_k$ is separated from $G_i$ and $G_j$ in Fig.~\ref{fig:IN_scheme})\cite{buldyrev2010catastrophic}.

Assuming that every node $v\in V$ represents a part of a larger (maybe continuous~\cite{Donges2012}) domain of interest, we assign it a weight $w_v>0$ which represents its share of the domain of interest. Therefore, when redefining any unweighted network measure, nodes should no longer be regarded as having a well-defined location, but their possible locations are constrained by the boundary of the represented area $w_v$. This implies that hypothetically splitting one node $v\in V_i$ into two adjacent and similar nodes $v'\in V_i$ and $v''\in V_i$ the former node weight $w_v$  should split up into $w_{v'}$ and $w_{v''}$ such that $w_v = w_{v'}+w_{v''}$. Since we assume links in a network to represent similarity or association between pairs of nodes, we treat $v'$ and $v''$ as being connected to all neighbours of the former node $v$ and to each other. Given that the sampling $G$ already provides a good approximation of the underlying complex system's structure, this node-splitting operation merely refines the network representation and should not affect the network measures much. Taking this into account yields a four-step construction mechanism for transforming unweighted network measures into their weighted counterparts as suggested in \cite{heitzig2012node}:
\begin{enumerate}
\item[(a)] Sum up weights $w_v$ whenever the unweighted measure counts nodes.
\item[(b)] Treat every node $v\in V$ as connected with itself.
\item[(c)] Allow equality for $v$ and $q$ wherever the original measure involves a sum over distinct nodes $v$ and $q$.
\item[(d)] ``Plug in'' n.s.i.\ versions of measures wherever they are used in the definition of other measures.
\end{enumerate}
(b) and (c) are derived from the fact that when splitting one node into two similar ones, the new nodes will be connected because of their similarity, (d) can be seen as an inductive step when applying (a) - (c) to an unweighted network measure. 
Treating every node as connected with itself, we additionally introduce the extended adjacency matrix $A^+_{pq} = A_{pq} + \delta_{pq}$, where $\delta_{pq}$ is Kronecker's delta.

\section{n.s.i.\ cross-network measures}

Here, we derive two informative measures for analysing networks of networks with heterogeneous node weights: (i)~n.s.i.\ cross-degree and (ii)~local n.s.i.\ cross-clustering coefficient. Further metrics can be easily derived analogously (see~\cite{heitzig2012node}). 

\subsection{n.s.i.\ cross-degree}

The unweighted cross-degree~\cite{donges2011investigating}
\begin{align}
    k_v^j = \sum_{q\in V_j} A_{vq}.
\end{align} 
measures the number of nodes $q\in V_j$ that are connected to a node $v\in V_i$ (typically $i \neq j$). Here we use mechanisms (a) and (b) to construct the n.s.i.\ cross-degree as
\begin{align}
    k_v^{j*} = \sum_{q\in V_j} w_q A^+_{vq}.
\end{align}
The n.s.i.\ cross-degree no longer just counts the number of nodes that $v$ is connected to. It rather measures the share of the whole sub-domain of interest given by the subnetwork $G_j$ that the node $v\in V_i$ is connected to. While the standard cross-degree can only take integer values in the range of $0,...,N-1$, the n.s.i.\ cross-degree $k_v^{j*}$ can assume real numbers in the range of $0\leq k_v^{j*}\leq W_j$, where $W_j = \sum_{q\in V_j}w_v$ is the total weight of all nodes $q$ in $V_j$, e.g., the area on the whole sub-domain of interest represented by the network $G_j$ (Fig.~\ref{fig:IN_scheme}).

\subsection{n.s.i.\ local cross-clustering coefficient}

The unweighted local cross-clustering coefficient~\cite{donges2011investigating}
\begin{align}
C_v^j = \frac{1}{k_v^j(k_v^j-1)} \sum_{p\neq q\in V_j} A_{vp} A_{pq} A_{qv}.
\end{align}
quantifies the probability that two randomly drawn neighbours $p,q\in V_j$ of $v\in V_i$ are also linked. Illustrating the construction mechanism, we convert the local cross-clustering coefficient into a n.s.i.\ cross-measure step by step:
\begin{align}
C_v^{j} &\stackrel{(a)}{\rightarrow} \frac{1}{k_v^j(k_v^j-1)} \sum_{p\neq q\in V_j} A_{vp} w_p A_{pq} w_q A_{qv} \nonumber  \\
&\stackrel{(b)}{\rightarrow} \frac{1}{k_v^j(k_v^j-1)} \sum_{p\neq q\in V_j} A^+_{vp} w_p A^+_{pq} w_q A^+_{qv} \nonumber \\
&\stackrel{(c)}{\rightarrow} \frac{1}{(k_v^j)^2} \sum_{p,q\in V_j} A^+_{vp} w_p A^+_{pq} w_q A^+_{qv} \nonumber \\
&\stackrel{(d)}{\rightarrow} \frac{1}{(k_v^{j*})^2} \sum_{p,q\in V_j} A^+_{vp} w_p A^+_{pq} w_q A^+_{qv} = C_v^{j*} \in [0,1]. \label{eq:clustering}
\end{align}
This only works if $k_v^{j*}>0$. Otherwise one may consider $C_v^{j*}$ as undefined or set $C_v^{j*}=0$. The n.s.i.\ local cross-clustering coefficient gives the probability that two randomly chosen points on the continuous domain of interest of the subnetwork $G_j$ which are neighbours of a point in the area represented by the node $v\in V_i$ are also neighbours. 

\section{Application 1: Spatial network model}

\begin{figure}[t]   
    \centering
    \includegraphics[width = 0.9\columnwidth]{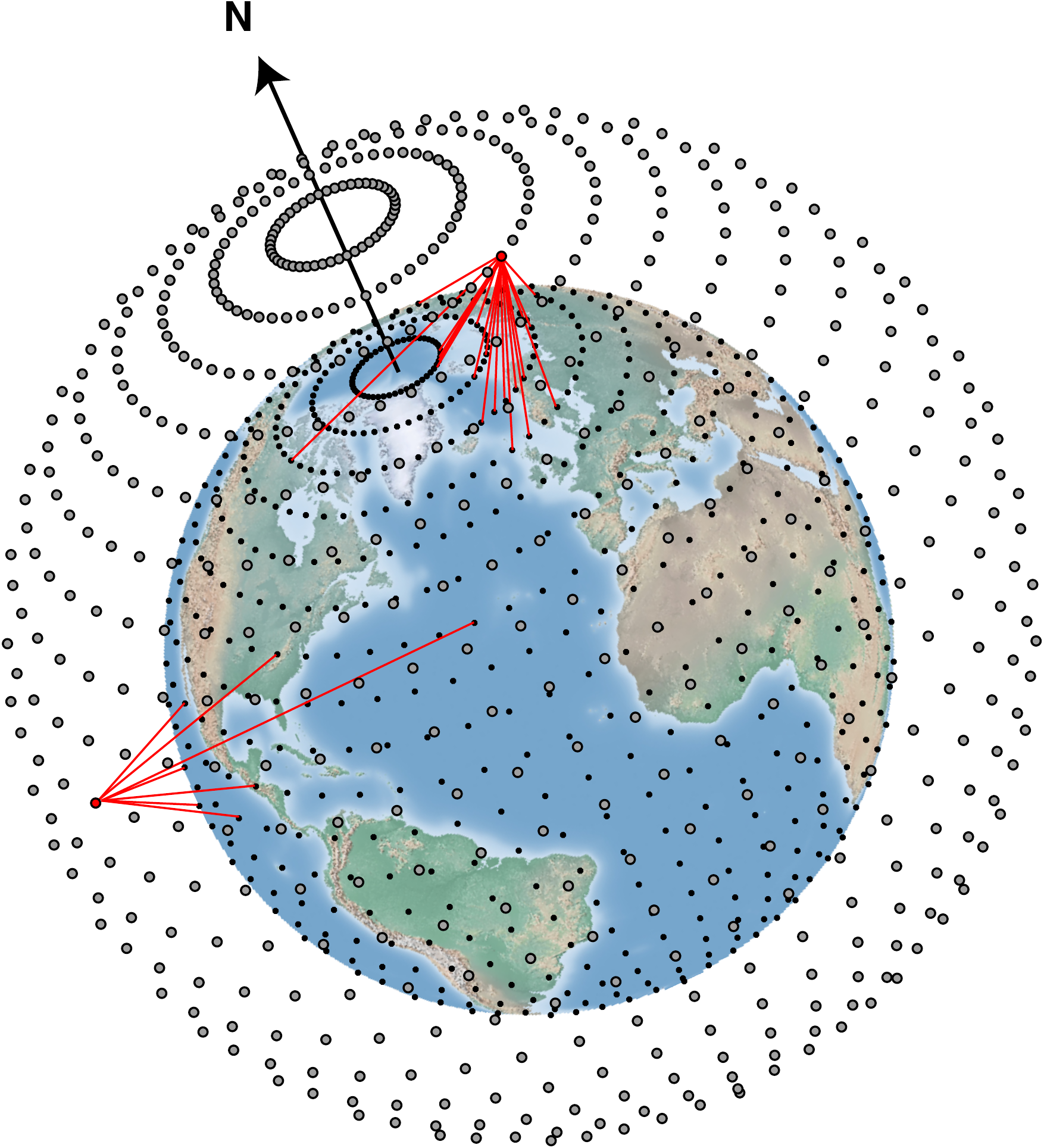} 
    \caption{(Colour online) Model of a spatial network $G=(V,E)$ consisting of two subnetworks $G_1=(V_1,E_{11})$ (filled circles) and $G_2=(V_2, E_{22})$ (open circles) representing layers parallel to the Earth's surface and links between nodes put randomly according to the distance between them. Only cross-links (red solid lines) emerging from two nodes in the upper layer are shown.
    }
    \label{fig:test_scheme}
\end{figure}

Assume a global model network $G=(V,E)$ consisting of two subnetworks $G_1=(V_1,E_{11})$ and $G_2=(V_2,E_{22})$ representing layers parallel to the Earth's surface with a distance $d \ll R$, where $R$ is the Earth's mean radius, as a typical representative of coupled climate networks \cite{donges2011investigating,schultz2010coupled} and other spatial networks~\cite{Barthelemy2011spatial} (Fig.~\ref{fig:test_scheme}). The node distribution in each layer is chosen to be identical and nodes are distributed globally with angular distance of $2.5^\circ$ in latitude and $5.0^\circ$ in longitude yielding a total number of $N = 10,658$ nodes. 
Links between nodes are considered to display a certain degree of similarity meaning that geographically close nodes are more likely to be linked than those spanning a larger distance. However, a small number of long-distance links such as teleconnections in climate networks \cite{tsonis2006networks} or intercontinental flights in airline networks~\cite{Gastner2006} may also be present. Therefore, in our model, links between all pairs of nodes $v$ and $q$ are introduced with a probability $P(s_{vq}) \propto \exp\left(-s_{vq} / \lambda \right)$, where $s_{vq}$ is the geodesic distance between $v$ and $q$. The typical length scale $\lambda=1000$\ km was chosen to yield a link density of $\rho \approx 0.05$. Regions close to the poles are expected to display larger local link density due to the increased node density in these areas (Fig.~\ref{fig:test_scheme}). The given setting provides an idealisation of typical spatially embedded networks~\cite{Barthelemy2011spatial}, where the nodes are located on a geographical grid having not necessarily a homogeneous node density. In this case, nodes $v$ at different latitudes $\theta_v$ represent a different share of the Earth's surface. We assign an individual weight $w_v = \cos(\theta_v)$ to every node $v$ so that nodes close to the poles have less weight than nodes near the equator. This choice of weight approximates the area actually represented by a node in an angularly regular spherical grid~\cite{heitzig2012node,tsonis2006networks}.

\begin{figure}[t] 
    \centering   
    \subfigure{\includegraphics[width = 0.9\columnwidth]{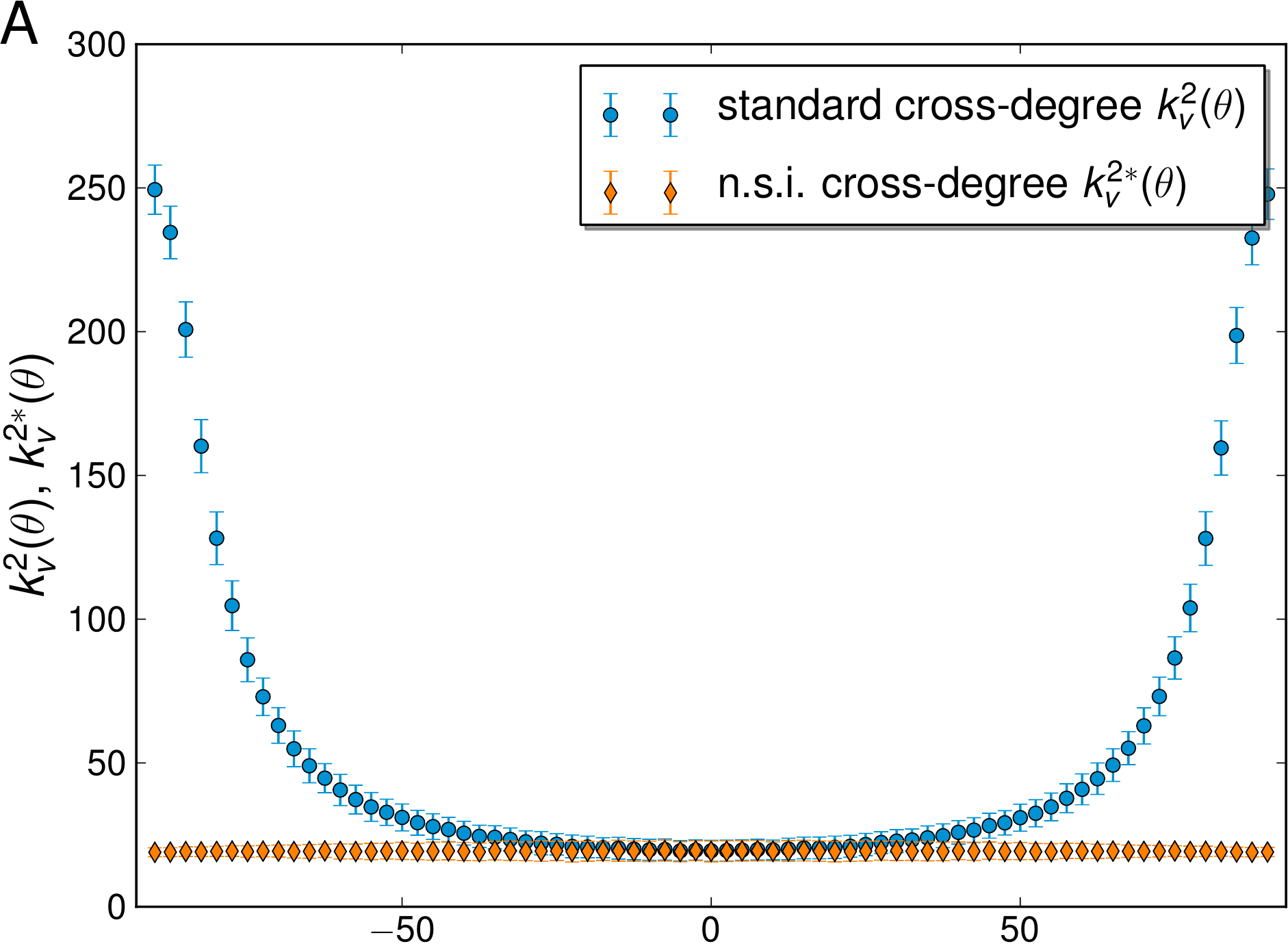}}
    \subfigure{\includegraphics[width = 0.9\columnwidth]{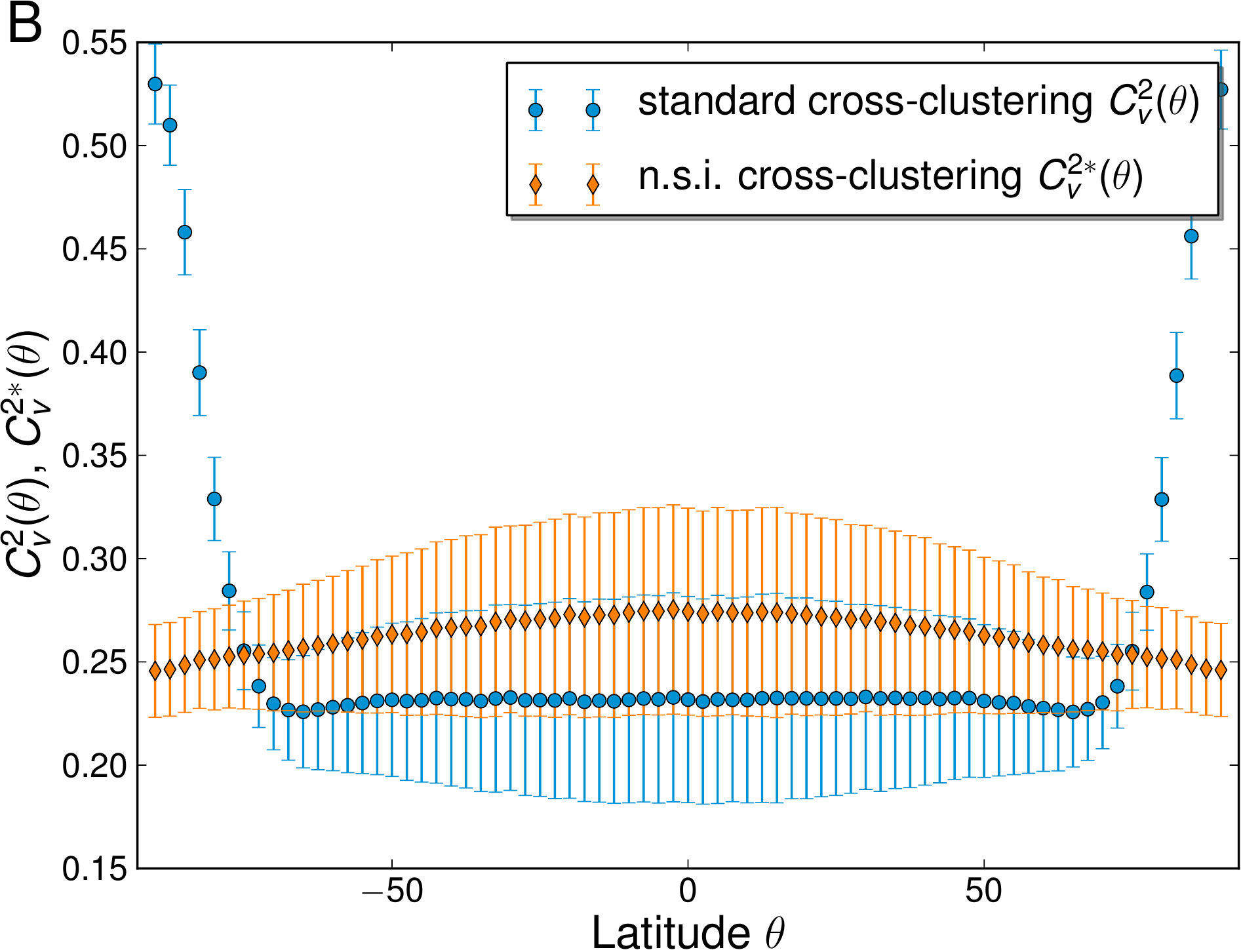}}
	\caption{(Colour online) Zonal averages of local standard cross-network measures and the corresponding n.s.i.\ cross-network measures for a spatial network model. (A) N.s.i.\ cross-degree $k_v^{2*}(\theta)$ and (B) n.s.i.\ local cross-clustering coefficient $C_v^{2*}(\theta)$ provide an improved representation of the underling model system's properties when approaching the poles ($\theta\rightarrow \pm 90^\circ$). Symbols represent means, error bars standard deviations based on an ensemble of $100$ realisations of the network model.
    }
    \label{fig:test_nsi_measures}
\end{figure}

In analogy to studies showing that the geometrical structure of a spatially embedded network strongly influences the distribution of network measures \cite{Henderson2011geometric,Rheinwalt2012}, we expect that local network properties, which are mainly based on counting adjacent nodes, should show remarkably increased values in regions with high node density. We compute zonal averages for the standard cross-degree $k_v^{2}(\theta)$ and the n.s.i.\ cross-degree $k_v^{2*}(\theta)$ for all nodes at latitude $\theta$ (only the results for $G_1$ are shown due to symmetry). The values of $k_v^{2}(\theta)$ rapidly increase towards the poles (Fig.~\ref{fig:test_nsi_measures}A), while due to the inhomogeneous node density, values are low at the equator. In comparison, $k_v^{2*}(\theta)$ which measures the area on the surface of $G_2$ that is connected to $v$ exhibits a constant value even close to the poles (Fig.~\ref{fig:test_nsi_measures}A). Here, the increased node density in this area compensates with the comparatively small node weight. Since we have considered a homogeneous network with no nodes having any outstanding properties, but pairs of nodes being randomly connected according to their distance, $k_v^{2*}(\theta)$ represents the network's topology much better than $k_v^{2}(\theta)$.

The behaviour of the standard local cross-clustering coefficient $C_v^{2}(\theta)$ is similar to that of $k_v^{2}(\theta)$ and the values of $C_v^{2}(\theta)$ increase rapidly towards the poles (Fig.~\ref{fig:test_nsi_measures}B). Due to the increasing node and local link density at the poles, nodes in this area have far more neighbours in their closer geographical surrounding than nodes near the equator. Since those neighbours are also more likely to be found close to the pole, it is quite probable that two neighbours of a node near a pole are also neighbours, yielding higher values of $C_v^{2}(\theta)$. $C_v^{2*}(\theta)$ is more robust to these circumstances showing an almost constant distribution of values over the whole globe. However, $C_v^{2}(\theta)$ and $C_v^{2*}(\theta)$ do not converge in the equatorial region. This disparity can be dealt with by further adjusting the n.s.i.\ cross-network measures and introducing a typical weight $\omega$ in the definition of the measures~\cite{heitzig2012node}.

\section{Application 2: International trade network}

Trade relationships between actors in world trade have recently been intensively studied \cite{garlaschelli2005structure, hidalgo2007product, goswami2012} and, in a globalised world, provide a particularly interesting example of a complex spatial network of interacting networks. The following analysis of a trade network illustrates potential for applications of interacting network measures pointing out the importance of taking into account properly chosen node weights $w_v$. In contrast to the above application, where node weights reflected the surface area represented by nodes, here we are interested in node weights indicating the national economic power that is involved in trade relationships.

\begin{figure}
    \centering
    \includegraphics[width=0.9\columnwidth]{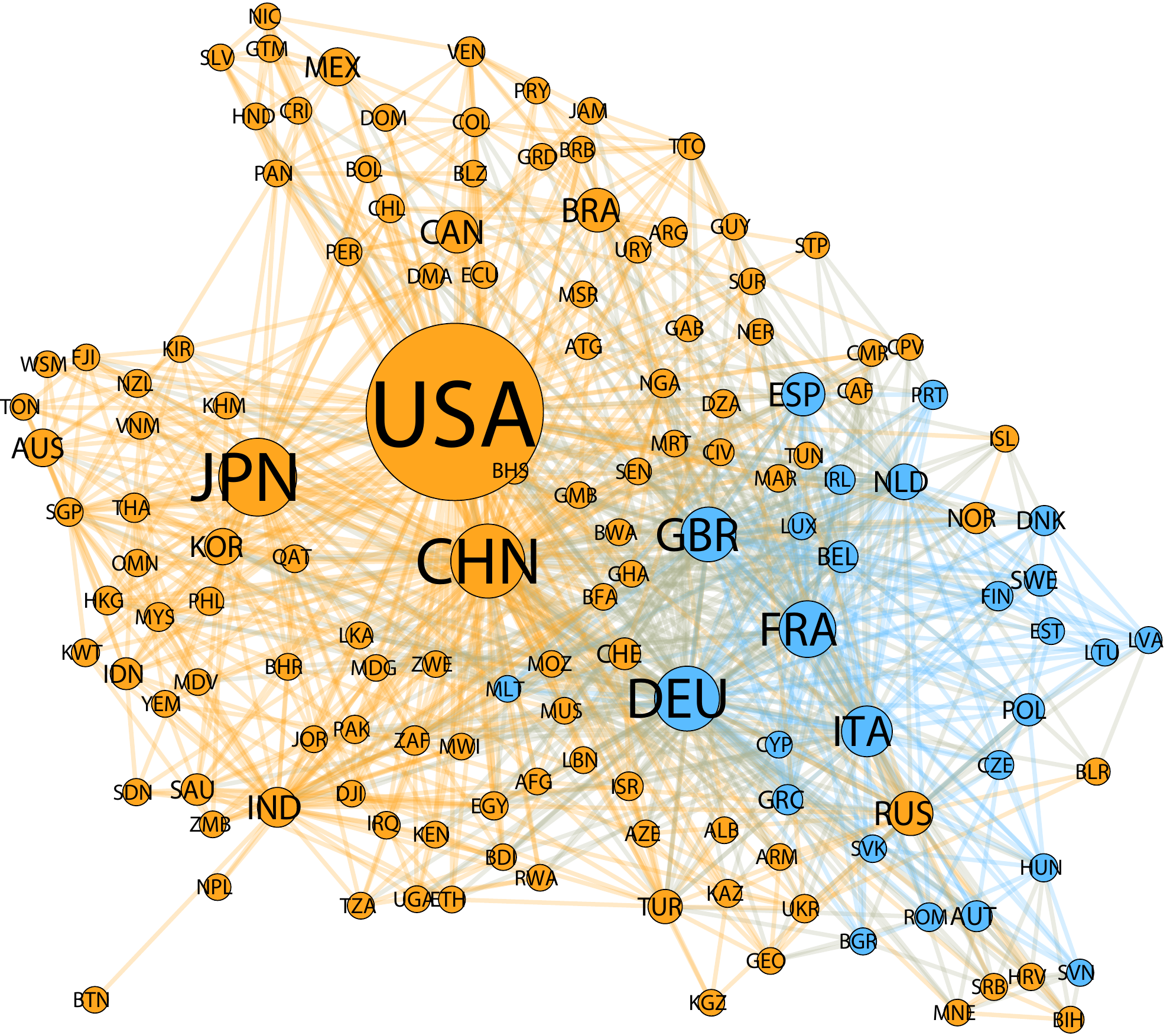}
    \caption{(Colour online) Visualisation of a trade network consisting of 142 countries that divides into two subnetworks, one representing the 27 members of the EU (blue) and one representing all non-EU countries (orange) using the force atlas 2 algorithm in \texttt{Gephi}\cite{bastian2009gephi}. Node size is proportional to node weight $w_v$: the countries' GDP in 2008.}
    \label{fig:tradenetwork}
\end{figure}

In particular, we perform an analysis of trading structure across the border of the common market of the European Union (EU) by studying a network of networks consisting of two subnetworks: $G_1$ representing all 27 EU countries and $G_2$ containing 115 non-EU countries (Fig.~\ref{fig:tradenetwork}). Nodes $v$ represent countries and links indicate a significant amount of trade between countries (as in \cite{baskaran2011}). Specifically, a link is created if the total value of the mutual reported trade in 2009 according to comtrade.un.org between two countries accounts for at least 5\ \% of the total trade of any of the two countries. Visualising this trade network indicates that defining EU and non-EU countries as two subnetworks is justified from the data by the fact that the nodes representing EU countries tend to form a tight cluster~\cite{donges2011investigating} (Fig.~\ref{fig:tradenetwork}). Additionally, we assign a weight $w_v$ to every node according to the 2008 Gross Domestic Product (GDP in USD). The GDP data is provided by the International Monetary Fund (IMF, http://www.imf.org). By doing so we do not only consider the total number of countries participating in trading but also the strength of different national economies.

\begin{figure}
    \centering
    \subfigure{\includegraphics[width = 0.8\linewidth]{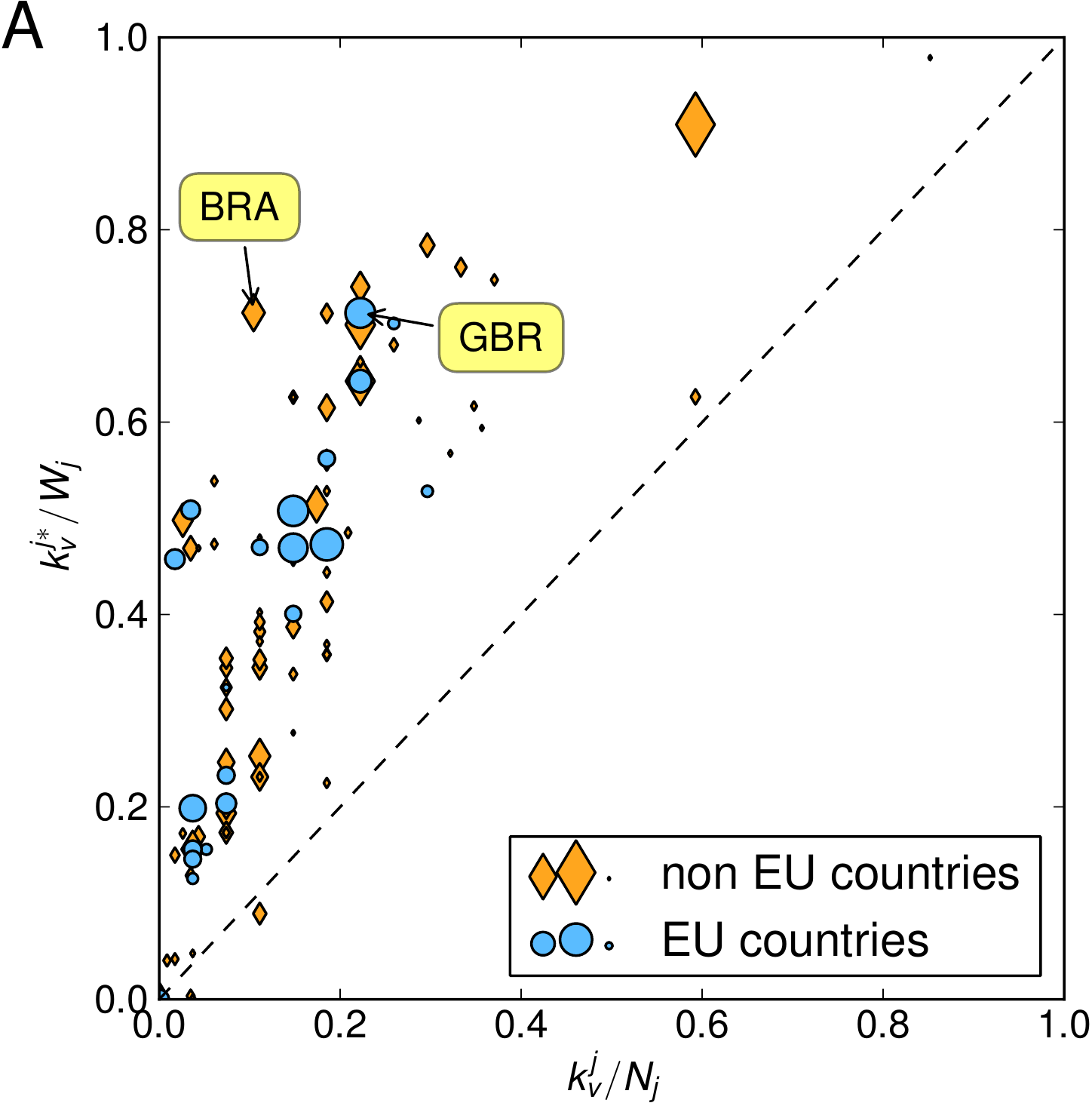}}
    \subfigure{\includegraphics[width = 0.8\linewidth]{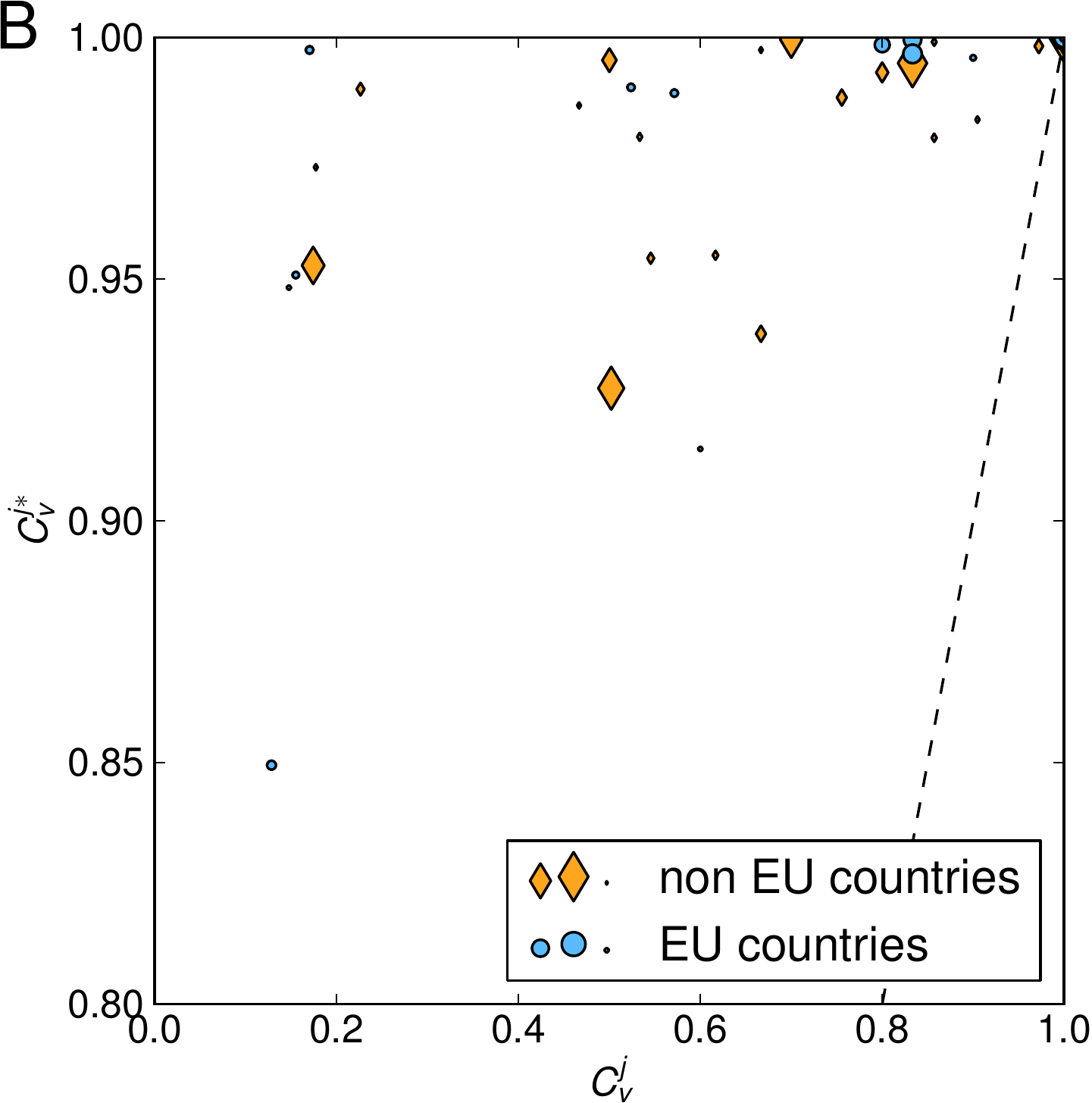}}
    \caption{(Colour online) Scatter plots of (A) normalised n.s.i.\ cross-degree $k_v^{j*}/W_j$ and (B) n.s.i.\ local cross-clustering coefficient $C_v^{j*}$ with their unweighted counterparts for the network of significant trade relations in 2009. $C_v^{j*}$ is shown for all nodes with $k_v^j > 1$. Symbol size is proportional to node weight $w_v$: the countries' GDP in 2008. For reference, dashed lines indicate equality of normalised weighted and unweighted measures.
    }
    \label{fig:trade_results}
\end{figure}

We compute the n.s.i.\ cross-degree $k_v^{2*}$ as well as the n.s.i.\ cross-clustering coefficient $C_v^{2*}$ for $v\in V_1$ and vice versa (Fig.~\ref{fig:trade_results}). The dashed line indicates the expected values if $w_v$ would not correct the examined network measure at all. For the cross-degree as well as the local cross-clustering coefficient, we observe no linear dependency between the weighted and the unweighted network measures. However, for almost every node the values of n.s.i.\ cross-network measures are significantly larger than their unweighted analogues. This observation indicates that the unweighted measures underestimate the corresponding properties of the underlying trade structure (the domain of interest).

The cross-degree $k_v^j$ counts the number of significant trading partners of a country $v$ across the border of the common market. In contrast, $k_v^{j*}$ takes into account each country's GDP, meaning its value displays the total economic strength that $v$ is trading with (Fig.~\ref{fig:trade_results}A). This way we identify the two countries with the highest deviation from the equality $k_v^{j*} / W_j=k_v^{j} / N_j$ as Great Britain (GBR) for the EU countries and Brazil (BRA) for the non-EU countries. For both countries, $k_v^{j*} / W_j$ is much larger than $k_v^{j} / N_j$, implying that both have comparatively few trading partners across the border of the EU common market, but the GDP of those partners sums up to a large economic strength that BRA and GBR are trading with. This can indicate both, an either positive or negative trade balance, of the country with its partners across the EU boundary. BRA exports comparatively many goods to the EU countries with strong economies, whereas GBR mainly imports goods from strong economies across the boundary of the common market. In order to obtain a broad insight into the trading behaviour of a country it is therefore necessary to evaluate both versions of cross-degree as they may be interpreted as different properties.

The local cross-clustering coefficient $C_v^j$ gives the probability for a country $v\in V_i$ to form a triangle with countries $p,q\in V_j$. In terms of trading this implies that a considered economy $v\in V_i$ has, besides trading directly with an economy $q\in  V_j$, a probability of $C_v^{j*}$ for trading with $q\in  V_j$ via $p\in V_j$, where $p$ might serve as a middleman. Again $C_v^j$ only counts the total number of nodes, e.g., triangles, that a node $v\in V_i$ is connected to, whereas $C_v^{j*}$ corrects the estimation by taking the total GDP of each country into account. We obtain further corrections away from linear dependency with the majority of countries showing a n.s.i.\ local cross-clustering coefficient of $C_v^{j*} > 0.9$~(Fig.~\ref{fig:trade_results}B). This result can be understood by the fact that the value of $C_v^{j*}$ is dominated by nodes with large node weights~(Eq.~(\ref{eq:clustering})). Furthermore, in terms of trading it is plausible that a country with a large GDP, i.e.,\ high $w_v$, tends to trade with a lot of other countries around the globe. Therefore, the probability to form a cluster including at least one country of strong GDP is comparatively high, but only those clusters significantly affect $C_v^{j*}$. Clusters including two partners $p,q\in V_j$ with small $w_p$ and $w_q$ do not significantly contribute to the value of $C_v^{j*}$ as both weights are multiplied under the sum~(Eq.~(\ref{eq:clustering})). Furthermore, as we treat a link not only as a single trade relation between countries but rather as an approximation of a bundle of links connecting distinct actors in trading within the countries, it is clear that the properties of the whole bundle are hardly affected by the failure of a single trade connection. Since $C_v^{j*}$ mirrors this property, we again consider $C_v^{j*}$ an improved representation of the underlying trade system's properties as compared to $C_v^{j}$. We conclude that the EU-non EU cross-boundary trade network can be considered as being very robust against the failure of links in the network as all nodes tend to cluster strongly with their neighbours. As in the first application, by introducing a properly chosen typical weight $\omega$ in the definition of $C_v^{j*}$, further refinements of the measure are possible\cite{heitzig2012node}.

\section{Conclusion}

In this work, we have used a general method for transforming standard network measures into their weighted counterparts in order to develop node-splitting invariant cross-network measures. The latter provide an efficient and universal tool for analysing the interaction structure between subnetworks in a network of networks with inhomogeneous node weights rendering those measures more appropriate to study real-world problems.

We have proceeded in three steps: (i)~We have carried out the derivations for two local cross-network measures, n.s.i.\ cross-degree and n.s.i.\ local cross-clustering coefficient. It is important to emphasise that all network measures (local and global) can be transformed into their node-weighted correspondents as well \cite{heitzig2012node,donges2011investigating}. (ii)~Considering a prototypical spatial model network consisting of two interdependent subnetworks embedded on surfaces parallel to the Earth's surface, we have validated the behaviour of our two measures of choice. Both n.s.i.\ cross-degree as well as n.s.i.\ local cross-clustering coefficient have been shown to provide an improved representation of the model system's underlying network topology as compared to their unweighted analogues. (iii)~In order to illustrate the variety of possible applications of our approach, we have carried out an analysis of the complex network structure of EU -- non EU cross-boundary trade. It was shown that node-weighted cross-degree yields additional insights into the trade relations as it identifies countries showing strong economic dependencies on overseas trade such as Great Britain as well as countries driving the economy of the EU by exporting to countries with a strong economy such as Brazil. In addition, we find that the n.s.i.\ local cross-clustering coefficient indicates a high robustness of the trade network to partial failure of links in the network, e.g., the breaking of single trade relationships between actors in different countries.

Both applications point out that the n.s.i.\ cross-network measures provide a general tool for analysing networks of networks with heterogeneously distributed node importance, size or weight that appear in research fields as diverse as climatology, neuroscience or economics. These network measures enable us to provide an improved understanding of the underlying complex system as long as weights are chosen appropriately to reflect the share of the domain of interest that is represented by the nodes. 

\acknowledgments

This work has been financially supported by the Leibniz association (project ECONS) and the German National Academic Foundation. We thank R.V. Donner for useful comments as well as R. Grzondziel and C. Linstead for help with the IBM iDataPlex Cluster at the Potsdam Institute for Climate Impact Research (PIK). Network measures were computed using the Python package \texttt{pyunicorn} developed at PIK.

\bibliography{eplbib}{}
\bibliographystyle{eplbib}

\end{document}